\documentclass[
twocolumn,
letter]{jpsj3}
\usepackage{txfonts}
\usepackage{cite,overcite}
\usepackage{amsmath,amssymb,bm
}
\usepackage[dvipdfmx]{graphicx}
\usepackage{latexsym}
\usepackage{braket}
\usepackage{mathrsfs}
\usepackage{cancel}
\usepackage{color}
\usepackage{multirow}
\usepackage{mathtools}
\usepackage{mathrsfs}
\usepackage{comment}
\usepackage{cancel}
\renewcommand{\d}{{\rm d}}
\renewcommand{\i}{{\rm i}}
\newcommand{\e}{{\rm e}}

\renewcommand{\Re}{{\rm Re}\;}
\renewcommand{\Im}{{\rm Im}\;}
\usepackage{xcolor}

\usepackage{appendix}
\title{Relationship between the Electronic Polarization and the Winding Number in Non-Hermitian Systems}

\author{Shohei Masuda and Masaaki Nakamura}

\inst{Department of Physics, Ehime University Bunkyo-cho 2-5, Matsuyama, Ehime 790-8577}

\abst{
 We discuss an extension of the Resta's electronic polarization to
 non-Hermitian systems with periodic boundary conditions. We introduce
 the ``electronic polarization'' as an expectation value of the
 exponential of the position operator in terms of the biorthogonal
 basis. We found that there appears a finite region where the
 polarization is zero between two topologically distinguished regions,
 and there is one-to-one correspondence between the polarization and the
 winding number which takes half-odd integers as well as integers.  We
 demonstrate this argument in the non-Hermitian Su-Schrieffer-Heeger
 model.
}

\begin{document}

\maketitle

%\section{introduction}
%--Introduction--

So far, non-Hermitian quantum mechanics has been discussed for some
specific situations
\cite{Feshbach,Hatano-N1996,Hatano-N1997,Hatano-N1998,Bender-B,Bender-B-J}.
In recent years, this field has attracted much attention in connection
with topological phases \cite{Ashida-G-U,Bergholtz-B-K} due to its
relevance to the open quantum systems
\cite{Ozawa,Lu,El-Ganainy,Longhi}. The application of topology in
conventional Hermitian quantum systems has relied on the principle of
bulk-edge (-boundary) correspondence \cite{Wilczek-Zee2008,Schnyder-R-F-L2008,Ryu-S-F-L2010} where the edge
or surface states in open boundary systems can be characterized by
topological invariants calculated under periodic boundary
conditions (PBCs). However, it has been pointed out that such a bulk-edge
correspondence may not be established in non-Hermitian systems
\cite{Esaki-S-H-K2011,Lee2016,Shen-Z-F2018}.  As a result, a macroscopic
number of bulk states localize at a boundary of the system which is
called as the non-Hermitian skin effect.
The skin effect is topologically characterized by the winding numbers
defined by the non-Bloch band theory
%in the generalized Brillouin zone
\cite{Yao-W2018,Yokomizo-M2019,Kawabata-O-S}, and that defined by the
Bloch Hamiltonian and the complex spectrum in open systems
\cite{Gong-A-K-T-H-U,Kawabata-S-U-S,Okuma-K-S-S,Vecsei-D-N-S,Li-Mu-L-G},
the biorthogonal polarization
\cite{Kunst-E-B-B2018,Edvardsson-K-Y-B2020}, and so on.
%
%Therefore, the bulk-edge correspondence in non-Hermitian systems has
%been mainly discussed comparing the edge modes in open boundary
%conditions with topological numbers
%
% including winding numbers in
%several definitions \cite{Esaki-S-H-K2011,Lee2016,Leykam-B-H-C-N2017,
%Shen-Z-F2018,Yao-W2018,Yokomizo-M2019}.

In this paper, inspired by these studies, we discuss the electronic
polarization in non-Hermitian systems.  Throughout of this paper, we
consider only periodic systems, since we need to know fundamental
relationship between topological quantities, before proceeding to deal
with open boundary systems.
For conventional Hermitian quantum lattice systems, the electronic
polarization has been introduced by Resta to evaluate the expectation
value of the position operator in periodic systems as
$z^{(q)}=\braket{\Psi_0|\e^{\i(2q\pi/L)\hat{X}}|\Psi_0}$ with $q=1$,
where $\hat{X}=\sum_{j=1}^{L} \hat{x}_j$ with $\hat{x}_j$ being the
position operator at $j$-th site, $L$ is the number of lattices, and
$\ket{\Psi_0}$ is the ground-state many-body wave function
\cite{Resta1994,Resta1998}.  This quantity can also be interpreted as
the overlap between the ground state and a variational excited state
appearing in the
% argument of the 
Lieb-Schultz-Mattis (LSM) theorem
\cite{Lieb-S-M1961,Affleck-L,Affleck}. $z^{(q)}$ with $q\geq 2$ can be
interpreted as extensions of the LSM theorem for general magnetizations
and fillings \cite{Oshikawa-Y-A,Yamanaka-O-A}, and by later equivalent
discussions \cite{Aligia-O1999}.  Resta related $z^{(1)}$ with the
electronic polarization as $\lim_{L\to\infty}(e/2\pi)\Im\!\ln z^{(1)}$.
Here and hereafter we call $z^{(q)}$ itself ``polarization''.  The
polarization $z^{(q)}$ has been calculated in various systems
\cite{Resta-S1999,Aligia-O1999,Nakamura-V2002,Nakamura-T2002} and is
shown to characterize topological phases and topological transitions in
one-dimensional systems.  Recently, an extension of the polarization to
higher-dimensional systems has been proposed based on spiral boundary
conditions \cite{Nakamura-M-N}.

As such a fundamental relation, in this paper, we will point out that
the electronic polarization in non-Hermitian systems can be related to a 
winding number which may take half integers
\cite{Lee2016,Yin-J-L-L-C2018,Imura-T,Imura-T2020}.  We will demonstrate
this argument in the non-Hermitian Su-Schrieffer-Heeger (SSH) model
\cite{Su-S-H1979,Su-S-H1980,Yao-W2018}.

%This paper is organized as follows: In Sec.\,\ref{sec:2}, we introduce
%the polarization and the generalized winding number. In
%Sec.\,\ref{sec:3}, we discuss the relationship between polarization and
%generalized winding number through the analysis of the non-Hermitian SSH
%model. Finally, a summary and discussion are given in Sec.\,\ref{sec:4}.

%\section{Polarization and winding number} \label{sec:2}
We introduce the biorthogonal system. The right and left eigenstates for a 1D quadratic non-Hermitian
Hamiltonian in real space satisfy the following eigenvalue equations:
\begin{subequations}
\begin{align}
 \sum_{j'}H^{\mathstrut}_{jj'}\ket{\psi^{j',{\rm R}}_{k\mu}}
 =E^{\mathstrut}_{k\mu}\ket{\psi^{j,{\rm R}}_{k\mu}},\\
 \sum_{j'}H^{\dag}_{jj'}\ket{\psi^{j',{\rm L}}_{k\mu}}
 =E^{\ast}_{k\mu}\ket{\psi^{j,{\rm L}}_{k\mu}},
\end{align}
\label{Sch_eqns}
\end{subequations}
respectively, where
$\ket{\psi^{j,\alpha}_{k\mu}}=\mathcal{U}^{-1}_{jk}\ket{u^{\alpha}_{k\mu}}\,
(\alpha = {\rm R}, {\rm L})$, $\mu$ is a band index,
$\mathcal{U}^{\mathstrut}_{kj}=\e^{-\i k x_j}/\sqrt{L}$, and
$\ket{u^{\alpha}_{k\mu}}$ with the normalization condition
$\braket{u^{\rm L}_{k\mu}|u^{\rm R}_{k\nu}}=\delta_{\mu\nu}$ is the
eigenstate of the non-Hermitian Bloch Hamiltonian $H(k)=\sum_{j
j'}\mathcal{U}^{\mathstrut}_{kj}H_{j j'}\mathcal{U}^{-1}_{j' k}$.

%\subsection{Electronic polarization}
Then we define the polarization in non-Hermitian systems as follows;
\begin{equation}
 z^{(q)}=\braket{\Psi_{\rm L}|U^{q}|\Psi_{\rm R}},\quad
  U=\exp\left[\,\i\frac{2\pi}{L}\sum_{j=1}^{L}\hat{x}_j\,\right],
  \label{z-def}
\end{equation}
where $\ket{\Psi_{\rm R(L)}}$ is the Slater determinant constructed by
the occupied right (left) eigenstate, and $\hat{x}_j=j n_j$, where $n_j$ is
the number operator of the electron at the $j$-th unit cell.
The definition (\ref{z-def}) for $q=1$ was first introduced by Lee, {\it
et al.} \cite{Lee-L-Y2020}. They obtained the right and left many-body
ground states by numerical diagonalization in real space and calculated
$\Im\ln z^{(1)}$, which takes $0$ or $\pi$.  In this approach, they could deal with small-size
systems, and the behavior of $z^{(q)}$ itself became unclear.

In order to improve this problem, we extend Resta's approach for
Hermitian systems \cite{Resta1998} to the polarization in non-Hermitian
systems, so that we rewrite Eq.\,\eqref{z-def} in terms of the right and
left one-particle eigenstates:
\begin{align}
z^{(q)}&={\rm det}\, S^{(q)},\\
 S^{(q)}_{\mu,\nu}(k_{s},k_{s'})
 &=\sum_{j=1}^{L}\braket{\psi^{j,{\rm L}}_{k_{s}\mu}|
 \e^{\i \frac{2q\pi}{L}x_j} |\psi^{j,{\rm R}}_{k_{s'}\nu}},
\end{align}
where $k_s = (2\pi/L)(s+\theta/2\pi)$ with
$s=0,1,\ldots,L-1$ for the boundary conditions
$\ket{\psi^{j+L,\alpha}_{k\mu}}
=\e^{\i\theta}\ket{\psi^{j,\alpha}_{k\mu}}$, and the determinant is
restricted to the occupied states. Then the determinant is rewritten as
a product of the biorthogonal overlaps:
\begin{equation}
 z^{(q)}=(-1)^{qN/L}\prod_{s=0}^{L-1}\,\prod_{\mu \in {\rm occupied}}
  \braket{u^{\rm L}_{k_{s+q}\mu}|u^{\rm R}_{k_{s}\mu}}, \label{reduced-z}
\end{equation}
where $N$ is the number of electrons. The factor $(-1)^{qN/L}$ stems
from the antisymmetry of the determinant \cite{Nakamura-M-N}.
Since Eq.\,(\ref{reduced-z}) is a simple product of overlaps between
Bloch states with the wave vectors that differ by $k_q$, $z^{(q)}$ can
be calculated for large system sizes, and the behavior of $z^{(q)}$ in
the thermodynamic limit will be clarified.

%\subsection{Generalized winding number}
We consider a two-band non-Hermitian system with chiral symmetry $\tau_3
H(k)\tau_3 =-H(k)$. In this case, the Hamiltonian is written as
\begin{equation}
H(k)=
\begin{pmatrix}
&h_{+}(k)\, \\
\,h_{-}(k) &
\end{pmatrix}. \label{chiral-Hamiltonian}
\end{equation}
Then we define the winding number as follows \cite{Wilczek-Zee2008,Schnyder-R-F-L2008,Ryu-S-F-L2010};
\begin{equation}
 \nu=\frac{\i}{2\pi}\int^{2\pi}_{0}\d k \, r^{-1}(k)
  \frac{\partial}{\partial k}r(k).
%  =\frac{\i}{2\pi}\oint\d q \, q^{-1}.
  \label{w-def}
\end{equation}
where $r(k)=h_{+}(k)/\sqrt{h_+(k) h_-(k)}$. For Hermitian systems, $r^{-1}$ becomes $r^\ast$, and $\nu$ can only
take $0$ or $1$. However, for non-Hermitian systems, the winding number
$\nu$ can be half-integers \cite{Yin-J-L-L-C2018,Imura-T,Imura-T2020}.

For the two-band model with the single occupied band, the phase of the
polarization is related to the winding number by
Eq.\,\eqref{reduced-z} in the $L\to\infty$ limit as follows;
\begin{equation}
 \ln z^{(1)}
 =\i \pi +\sum_{s=0}^{L-1}
 \ln\braket{u^{\rm L}_{k_s}|u^{\rm R}_{k_{s-1}}}
 \simeq\i(1\pm\nu)\pi,
 \label{relation}
\end{equation}
which is well-defined unless $\nu$ is half-odd integers, where the signs ``$\pm$'' are related to the choice of the eigenstates \cite{Ryu-S-F-L2010}.

%\section{non-Hermitian SSH model} \label{sec:3}

\begin{figure}[t]
\centering
\includegraphics[width=80mm,pagebox=cropbox,clip]{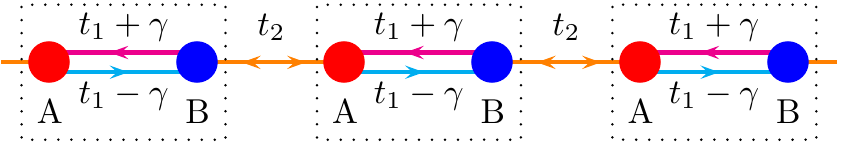}
\caption{(Color online) The non-Hermitian SSH model (\ref{SSH}). The dotted boxes
indicate the unit cells.}  \label{fig:model}
\end{figure}

\begin{figure}[t]
\centering
\includegraphics[width=80mm,pagebox=cropbox,clip]{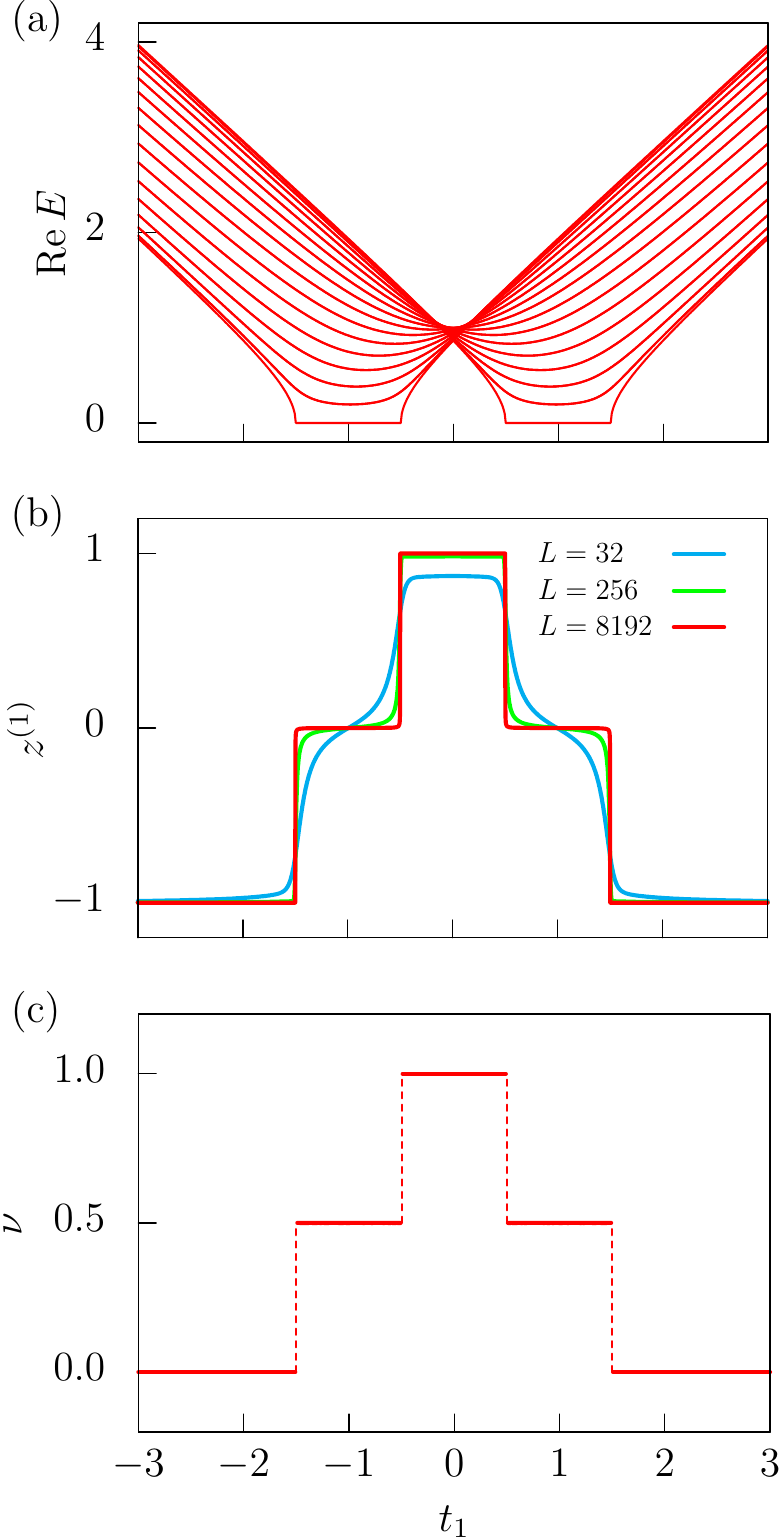}
\caption{(Color online) (a) The real part of energy eigenvalues, (b) the polarization
and (c) the winding number in the non-Hermitian SSH model
(\ref{SSH}) with $t_{2}=1$ and $\gamma=0.5$. The eigenvalues are
computed for $L=32$, and the polarization is calculated for $L=32,256$
and $8192$.}  \label{fig:result}
\makeatletter
\let\save@currentlabel\@currentlabel
\edef\@currentlabel{\save@currentlabel(a)}\label{fig:result1}
\edef\@currentlabel{\save@currentlabel(b)}\label{fig:result2}
\edef\@currentlabel{\save@currentlabel(c)}\label{fig:result3}
\makeatother
\end{figure}

%\subsection{model Hamiltonian}
We concider the non-Hermitian Su-Schrieffer-Heeger (SSH) model with
half-filling \cite{Su-S-H1979,Su-S-H1980,Yao-W2018} as illustrated in
Fig.\,\ref{fig:model}. The Hamiltonian in real space is given by
\begin{align}
 \mathcal{H}&=t_{1} \sum_{i=1}^{L}
 c^\dag_{i,{\rm A}} c^{\mathstrut}_{i,{\rm B}}
 +t_{2} \sum_{i=1}^{L} c^\dag_{i,{\rm B}} c^{\mathstrut}_{i+1,{\rm A}}
 + {\rm H.c.} \notag\\
 &\ + \gamma \sum_{i=1}^{L} c^\dag_{i,{\rm A}} c^{\mathstrut}_{i,{\rm B}}
 - {\rm H.c.},
 \label{SSH}
\end{align}
where $c^{\mathstrut}_{i,\alpha}$ $(c^\dag_{i,\alpha})$ denotes the
annihilation (creation) operator of a fermion in sublattice $\alpha$ at
the $i$-th unit cell. When $\gamma=0$, this model is Hermitian,
and phase transitions occur at $t_{1}=\pm t_{2}$. This Hamiltonian
can be written as the quadratic form:
\begin{equation}
 \mathcal{H}=\sum_{ij,\alpha\beta}
  c^\dag_{i,\alpha} H_{ij,\alpha\beta}c^{\mathstrut}_{j,\beta},
\end{equation}
so that it is diagonalized by the Fourier transformation
$c^{\mathstrut}_{k,\alpha}=\sum_{j}\mathcal{U}^{\mathstrut}_{kj}
c^{\mathstrut}_{j,\alpha}$ as follows:
\begin{align}
 \mathcal{H}
 &=\sum_{k,\alpha\beta}
 c_{k,\alpha}^\dag H_{\alpha\beta}(k) c_{k,\beta}^{\mathstrut},\\
 H(k)
 &=\sum_{\mu=1}^{2}(d_\mu + \i f_\mu)\tau_\mu, \label{k-Hamiltonian}
\end{align}
where 
\begin{equation}
\begin{array}{ll}
d_1= t_{1}+ t_{2} \cos k,\quad &f_1=0,\\
d_2=t_{2} \sin k,\quad &f_2 =\gamma,
\end{array}
\end{equation}
and $\tau_{\mu}\ (\mu=1,2,3)$ is Pauli matrices. This model has chiral
symmetry, so that the energy eigenvalues are symmetric about zero
energy. The square of the energy eigenvalue is given by
\begin{equation}
 E^2(k)
 =t_{1}^2+t_{2}^2-\gamma^2
 +2 t_{1}t_{2}\cos k
 +\i \,2t_{2}\gamma\sin k. \label{energy}
\end{equation}
The energy gap closes at the exceptional points (EPs) $(d_1,
d_2)=(-f_2,f_1)$ or $(f_2,-f_1)$, which leads to $t_{1}=-t_{2} \pm
\gamma$ for $k=0$ or $t_{1}=t_{2} \pm \gamma$ for $k=\pi$. For $k=0$ and
$\pi$, the energy eigenvalues are $E=\pm\sqrt{(t_{1}+t_{2})^2-\gamma^2}$
and $\pm\sqrt{(t_{1}-t_{2})^2-\gamma^2}$, respectively, so that they are
pure imaginaries when $t_{1}$ satisfies
$-t_{2}-\gamma<t_{1}<-t_{2}+\gamma$ or
$t_{2}-\gamma<t_{1}<t_{2}+\gamma$, respectively [see
Fig.\,\ref{fig:result1}].

%\subsection{Polarization and generalized winding number}
For the non-Hermitian SSH model in real space, we concider
the ``space inversion'' as $\mathcal{P}c_{j,{\rm
A(B)}}\mathcal{P}^{-1}=c_{L+1-j,{\rm B(A)}}$. Under this
transformation, the non-Hermitian Bloch Hamiltonian
\eqref{k-Hamiltonian} satisfies the following relation
\begin{equation}
\mathcal{P}H(k)\mathcal{P}^{-1}=H^\dag(k),
\end{equation}
%Actually, t
where $\mathcal{P}=\tau_1 \mathcal{K}$, and $\mathcal{K}$ is a conjugate
operator. Moreover, this transformation leads to
\begin{equation}
z^{(q)}
=\braket{\Psi_{\rm R}|\mathcal{P}U^{q}\mathcal{P}^{-1}|\Psi_{\rm L}}
=\e^{\i 2q \pi N/L}[z^{(q)}]^\dag. \label{z-z^dag relation}
\end{equation}
Since $N=L$ in the non-Hermitian SSH model at half-filling, we should
choose $q=1$. Then Eq.\,\eqref{reduced-z} is reduced to
\begin{equation}
 z^{(1)}=(-1)\prod_{s=0}^{L-1}\,
  \braket{u^{\rm L}_{k_{s+1}-}|u^{\rm R}_{k_{s}-}},
\end{equation}
where $\ket{u^{\rm R(L)}_{k_{s}-}}$ is the right (left) eigenstate of
the occupied band. For $t_{2}+\gamma<t_{1}$, the electrons of the system
are located on unit cells, so that we find that $z^{(1)}=-1$ for even
$L$ because $(2\pi/L)\sum_{x_j=1}^{L} x_j=\pi(L+1)$. Similarly, for
$-t_{2}+\gamma<t_{1}<t_{2}-\gamma$, the electrons of the system are
located on between a unit cell and the neighboring one, so that we find
that $z^{(1)}=1$ for even $L$ because of $(2\pi/L)\sum_{x_j=1}^{L}
(2x_j+1)/2=\pi(L+2)$. On the other hand, for
$-t_{2}-\gamma<t_{1}<-t_{2}+\gamma$ or
$t_{2}-\gamma<t_{1}<t_{2}+\gamma$, the electrons of the system are
located homogeneously, so that $z^{(1)}=0$ for even $L$ because
$(2\pi/L)\sum_{x_j=1}^{L} (2x_j+1/2)/2=\pi(2L+3)/2$. Here, we have used
the fact that $z^{(1)}$ is always real, since Eq.\,\eqref{z-z^dag
relation} leads to $z^{(1)}=[z^{(1)}]^\dag$ in the non-Hermitian SSH
model.

The numerical results of the polarization $z^{(1)}$ for different system
sizes are shown in Fig.\,\ref{fig:result2}, and $z^{(1)}$ distinguishes
three topologically distinct phases $z^{(1)}=-1,0,+1$, as expected.
Here, we have used anti-periodic boundary conditions
$c_{i+L,\alpha}=-c_{i,\alpha}$ to prevent divergences at EPs, at which
phase transitions occur. The results do not depend on
the boundary conditions except for the phase transition points. If the
regions with $\Re E=0$ are interpreted as gapless phases, the results
with $z^{(1)}=0$ are consistent with the LSM theorem. Note that
$z^{(1)}$ in a finite-size system has a finite value, so that
$(2\pi)^{-1}{\rm Im}\ln z^{(1)}$ is not a good quantity to determine
topologically different phases, especially for $z^{(1)}=0$ regions.

Using Eq.\,\eqref{w-def}, we obtain the following winding
number of the non-Hermitian SSH model:
\begin{equation}
 \nu=\int^{2\pi}_{0}\frac{\d k}{2\pi} \,
  \frac{ t_{2}(t_{1}\cos k
  +\i \gamma \sin k+t_{2})}{E^2},
\end{equation}
which is well defined except at EPs, where $E^2$ is given by
Eq.\,\eqref{energy}. The winding number is equal to the half of the
number of times that, in $d_1$-$d_2$ plane, EPs are surrounded by the
closed curve when $k$ ranges from $0$ to $2\pi$ \cite{Yin-J-L-L-C2018}.

The numerical results of the winding number $\nu$ are shown in
Fig.\,\ref{fig:result3}. Comparing them with the results of the
polarization $z^{(1)}$ in the thermodynamic limit ($L\to\infty$), we
find that there are one-to-one correspondences between $\nu=0,1/2,1$ and
$z^{(1)}=-1,0,+1$ for the topologically different phases. Therefore,
Eq.(\ref{relation}) becomes singular at $z^{(1)}=0$, and two quantities
in this case can be well related as $z^{(1)}=-\cos\nu\pi$.

\begin{figure}[t]
\centering
\includegraphics[width=80mm,pagebox=cropbox,clip]{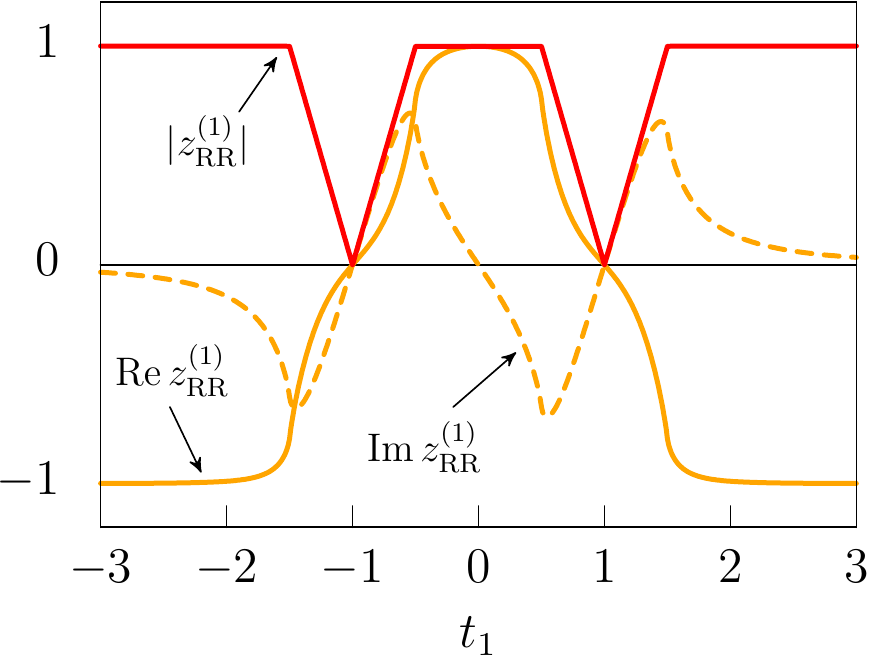}
\caption{(Color online) The real part (orange line), the imaginary part (orange dashed
line) and the absolute value (red line) of the polarization, which is
given by Eq.\,\eqref{reduced-z^(1)_RR}, in the non-Hermitian SSH model
(\ref{SSH}) with $t_{2}=1$ and $\gamma=0.5$ for $L=8192$.}
\label{fig:z-RR}
\end{figure}

% RR
For comparison to the polarization $z^{(1)}$, we concider the
polarization defined by the occupied right eigenstate. In the
non-Hermitian SSH model, this is given by
\begin{equation}
 z^{(1)}_{\rm RR}=(-1)\prod_{s=0}^{L-1}\,
  \braket{u^{\rm R}_{k_{s+1}-}|u^{\rm R}_{k_{s}-}},
  \label{reduced-z^(1)_RR}
\end{equation}
where $\ket{u^{\rm R}_{k\mu}}$ is normalized as $\braket{u^{\rm
R}_{k\mu}|u^{\rm R}_{k\mu}}=1$, and the numerical results are shown in Fig.\,\ref{fig:z-RR}. Unlike
$z^{(1)}$,
%defined in Eq.\,\eqref{z-def} with the relation in Eq.\,\eqref{z-z^dag
%relation}, 
$z^{(1)}_{\rm RR}$ is generally complex. We find that ${\rm
Re}\,z^{(1)}_{\rm RR}\neq 0$ in the region that ${\rm Re}\, E=0$ and
$\nu=1/2$.  Therefore, $z^{(1)}_{\rm RR}$ is not a good index to
characterize the different phases comparing with $z^{(1)}$. On the other
hand, $|z^{(1)}_{\rm RR}|$ shows liner behavior in terms of $t_1$,
%and $|z^{(1)}_{\rm RR}|=0$ at the point that $t_{1}=\pm t_{2}$, 
and $|z^{(1)}_{\rm RR}|\neq 1$ or $|z^{(1)}_{\rm RR}|=1$ correspond the
phases with $\nu=1/2$ and $\nu=0,1$, respectively. Thus $|z^{(1)}_{\rm
RR}|$ is better topological index than ${\rm Re}\,z^{(1)}_{\rm RR}$ and
${\rm Im}\,z^{(1)}_{\rm RR}$, in this case. Comparing the results of the
two-types of polarization $z$ and $z_{\rm RR}$ defined by
Eqs. (\ref{z-def}) and (\ref{reduced-z^(1)_RR}), it is sufficient to use the
former one given by a real number to describe the topological phases and
transitions in the present non-Hermitian system as discussed in
Ref.~\citen{Bender-B-J}.

%\section{Summary and Discussion} \label{sec:4}

In wummary, we have studied an extension of the Resta's electronic polarization
\cite{Resta1998} to non-Hermitian systems with chiral symmetry.  We have
given the reduction formalism for the polarization (\ref{reduced-z})
which enables us to deals with large systems close to the thermodynamic
limit. Then the relationship between the polarization and the winding
number has been discussed.

We have analyzed the non-Hermitian SSH model and shown numerically that
the polarization defined by the biorthogonal basis (\ref{z-def}) takes
three different values $z^{(1)}=-1,0,+1$ for the topologically distinct
phases in the thermodynamic limit.  We have also found that there is
one-to-one correspondence with the winding number
$\nu=0,1/2,1$ defined in Refs.~\citen{Lee2016,Yin-J-L-L-C2018}.
Furthermore, this relationship has been shown analytically as
Eq.~(\ref{relation}).  Here we should be careful to take the
thermodynamic limit of $z^{(1)}$: depending on how to take the
thermodynamic limit, the region with $\nu=1/2$ may be lost
\cite{Lee-L-Y2020}.

The topological phases for the non-Hermitian skin effect in open systems
have been discussed by a winding number in a different definition
\cite{Yao-W2018} from the present one for periodic systems. This winding number takes only two values, and the
``transition point'' where the winding number changes its value appears
to be different. However their difference is only whether the
integration is defined in the generalized Brillouin zone
or in the ordinary Brillouin zone \eqref{w-def}. Therefore the present
study may contribute to analyze the skin effect by generalizing the
polarization operator.
%The topological phases in non-Hermitian systems have also been discussed
%by a winding number in a different definition \cite{Yao-W2018} from the
%present one (see Appendix \ref{Appendix A}).  This winding number takes
%only two values and the ``transition point'' where the winding number
%changes its value appears to be different. Therefore the present study
%may contribute to define topological transitions in non-Hermitian
%systems in more precise way.
%
There is another topological number ``biorthogonal polarization''
defined in open boundary systems \cite{Kunst-E-B-B2018,Edvardsson-K-Y-B2020}.
This quantity is considered to characterize the number of edge modes
\cite{Edvardsson-K-Y-B2020}. Therefore, despite having a similar
appearance, there is no direct correspondence with the present
electronic polarization.

We have also examined the polarization defined only by right eigenstates
$z^{(1)}_{\rm RR}$. This quantity takes a complex value and is not
explicitly related to the winding number.  In contrast to this, the
polarization defined by the biorthogonal basis $z^{(1)}$ is a real
number \cite{Bender-B-J} and there is a clear one-to-one correspondence
with the winding number, so that $z^{(1)}$ is considered to be more
suitable definition than $z^{(1)}_{\rm RR}$ for the present closed system.

%%%%%%%%%%%%%%%%%%%%%%%%%%%%%%%%%%%%%%%%%%%%%%%%%%%%%%%%%%%%%%%%%%%%%%%
%% Acknowledgment
%%%%%%%%%%%%%%%%%%%%%%%%%%%%%%%%%%%%%%%%%%%%%%%%%%%%%%%%%%%%%%%%%%%%%%%
\section*{Acknowledgment}
The authors thank N.~Hatano and K.~Imura for valuable discussions.
M.~N. acknowledges the Visiting Researcher's Program of the Institute
for Solid State Physics, The University of Tokyo, and the research
fellow position of the Institute of Industrial Science, The University
of Tokyo. M.~N. is supported partly by MEXT/JSPS KAKENHI Grant
No.~JP20K03769.

%
%	reference
%

\end{document}